\documentclass[aps,prb,twocolumn,superscriptaddress,english]{revtex4-2}
\usepackage[utf8]{inputenc}

\usepackage{amsmath}
\usepackage{amssymb}
\usepackage{graphicx}
\usepackage{xcolor}
\usepackage{siunitx}
\usepackage{lipsum}
\usepackage{orcidlink}

\usepackage{lineno}

\usepackage{hyperref}
\usepackage{orcidlink}
\usepackage{xr-hyper}

\makeatletter
\def\@fnsymbol#1{\ensuremath{\ifcase#1\or \dagger\or *\or \ddagger\or
   \mathsection\or \mathparagraph\or \|\or **\or \dagger\dagger
   \or \ddagger\ddagger \else\@ctrerr\fi}}
\makeatother
\def\thefootnote{*}\footnotetext{These authors contributed equally to this work.}

\begin{document}

\author{Simone Di Cataldo\thefootnote \orcidlink{0000-0002-8902-0125}}\email{simone.dicataldo@uniroma1.it}
\affiliation{Dipartimento di Fisica, Sapienza Universit\`a di Roma, Piazzale Aldo Moro 5, 00187 Roma, Italy}
\affiliation{Instit\"{u}t f\"{u}r Festk\"{o}rperphysik, TU Wien, Wiedner Hauptstraße 8-10, Vienna 1050, Austria}
\author{Maria Rescigno\thefootnote \orcidlink{0000-0002-5976-4449}}\email{maria.rescigno@uniroma1.it}
\affiliation{Dipartimento di Fisica, Sapienza Universit\`a di Roma, Piazzale Aldo Moro 5, 00187 Roma, Italy}
\affiliation{Laboratory of Quantum Magnetism, Institute of Physics, \'{E}cole Polytechnique F\'{e}d\'{e}erale de Lausanne, CH-1015 Lausanne, Switzerland}
\author{Lorenzo Monacelli \orcidlink{0000-0002-6381-3741}}
\affiliation{Dipartimento di Fisica, Sapienza Universit\`a di Roma, Piazzale Aldo Moro 5, 00187 Roma, Italy}
\author{Umbertoluca Ranieri \orcidlink{0000-0003-2026-6132}}
\affiliation{Centre for Science at Extreme Conditions and School of Physics and Astronomy, University of Edinburgh, EH9 3FD Edinburgh, UK}
\author{Richard Gaal \orcidlink{0000-0001-5181-423X}}
\affiliation{Laboratory of Quantum Magnetism, Institute of Physics, \'{E}cole Polytechnique F\'{e}d\'{e}erale de Lausanne, CH-1015 Lausanne, Switzerland}
\author{Stefan Klotz \orcidlink{0000-0002-5682-4960}}
\affiliation{Sorbonne Université, UMR CNRS 7590, Institut de Minéralogie, de Physique des Matériaux et de Cosmochimie (IMPMC), 5 Place Jussieu,
75005 Paris, France}
\author{Jacques Ollivier}
\affiliation{Institut Laue-Langevin, 71 Avenue des Martyrs, Cedex 9, Grenoble, France}
\author{Michael Marek Koza \orcidlink{0000-0002-5133-8584}}
\affiliation{Institut Laue-Langevin, 71 Avenue des Martyrs, Cedex 9, Grenoble, France}
\author{Cristiano De Michele \orcidlink{0000-0002-8367-0610}}
\affiliation{Dipartimento di Fisica, Sapienza Universit\`a di Roma, Piazzale Aldo Moro 5, 00187 Roma, Italy} 
\author{Livia Eleonora Bove \orcidlink{0000-0003-1386-8207}}\email{livia.bove@upmc.fr} 
\affiliation{Dipartimento di Fisica, Sapienza Universit\`a di Roma, Piazzale Aldo Moro 5, 00187 Roma, Italy}
\affiliation{Laboratory of Quantum Magnetism, Institute of Physics, \'{E}cole Polytechnique F\'{e}d\'{e}erale de Lausanne, CH-1015 Lausanne, Switzerland}
\affiliation{Sorbonne Université, UMR CNRS 7590, Institut de Minéralogie, de Physique des Matériaux et de Cosmochimie (IMPMC), 5 Place Jussieu,
75005 Paris, France}

\title{Giant splitting of the hydrogen rotational eigenenergies in the C$_2$ filled ice}

\begin{abstract}
Hydrogen hydrates present a rich phase diagram influenced by both pressure and temperature, with the so-called C$_2$ phase emerging prominently above 2.5 GPa. In this phase, hydrogen molecules are densely packed within a cubic ice-like lattice and the interaction with the surrounding water molecules profoundly affects their quantum rotational dynamics. Herein, we delve into this intricate interplay by directly solving the Schr\"{o}dinger's equation for a quantum H$_2$ rotor in the C$_2$ crystal field at finite temperature, generated through Density Functional Theory. Our calculations reveal a giant energy splitting relative to the magnetic quantum number of $\pm$3.2 meV for $l=1$. Employing inelastic neutron scattering, we experimentally measure the energy levels of H$_2$ within the C$_2$ phase at 6.0 and 3.4 GPa and low temperatures, finding remarkable agreement with our theoretical predictions. These findings underscore the pivotal role of hydrogen--water interactions in dictating the rotational behavior of the hydrogen molecules within the C$_2$ phase and indicate heightened induced-dipole interactions compared to other hydrogen hydrates.
\end{abstract}

\maketitle

The phase behavior of hydrogen hydrates represents a captivating realm of inquiry at the intersection of physics and quantum chemistry, offering insights into the complex interplay of molecular interactions under extreme conditions and quantum motions under extreme confinements. At pressures above 0.1 GPa and below $\sim$0.5 GPa, mixtures of water and hydrogen molecules crystallize in a non-stoichiometric clathrate compound, which is characterized by polyhedral cavities formed by hydrogen-bonded water molecules encaging a variable amount of non-bonded H$_2$ molecules \cite{Mao_Science_2002_sII, Mao_PNAS_2004_sII, Lokshin_PRL_2004_sII_neutrons,Strobel_JPCC_2011_H2H2O, Bazarkina_Elements_2020_review}. The dynamics of the (freely rotating) hydrogen molecules within the cages stabilizes the highly porous ice structure \cite{Tse_PNAS_2003_clathrates}. At low temperatures, the quantum motion of the ``encaged" guest H$_2$ molecules has been characterized by means of inelastic neutron scattering and Raman spectroscopy experiments, mostly on samples recovered to ambient pressure \cite{Giannasi_JCP_2008, Ulivi_PRB_2007, Colognesi_JPCA_2013_H2_sII, Ranieri_JCP_2024}, and reveals to be similar to the one observed for H$_2$ trapped in other nanocavities \cite{Narehood_PRB_2002_H2_in_CN, Mondelo_JCP_2015_H2_in_SWNT, Fitzgerald_PRB_2010_H2_in_MOF, Strobel_PRL_2018_H2_organic, Horsewill2009, Horsewill_PRB_2012_H2_inelastic, Xu_PRL_2014_confirming}. 
The anisotropy in the hydrogen--water interaction potential has been shown to be effective in splitting the quantized translational and rotational energy levels; however, such splittings are rather small, indicating a weak guest--host interaction. 
Ba\v{c}i\'c and co-workers led the way in developing the theoretical framework for describing this cage potential, resulting in a more precise assessment of the energy levels that define the quantum dynamics of H$_2$. These computational studies shed light on a crucial aspect of dynamics: the interaction between translational and rotational angular momentum. This interaction is mediated by the 5D-cage potential, which includes terms influenced by the orientation of the axis of the H$_2$ molecule in relation to the normal direction of the cage wall \cite{Xu_JCP_128_2008, Xu_JCP_129_2008, Sebastianelli_JACS_2010, Xu_PRL_2014_confirming}.
The quantum dynamics of the H$_2$ molecule confined in a clathrate hydrate structure was also studied as a function of pressure up to 0.5 GPa \cite{Ranieri_JPCC_2019_quantumH2} showing only a small impact on the energy levels and splittings, despite the varying water-hydrogen distances with pressure. 

At pressures above $\sim$0.5 GPa, the hydrates of hydrogen adopt so-called \textit{filled ice} phases, in which the water sublattice is structurally similar to ice \cite{Amos_JPCL_2017_C0, DelRosso_NatComm_2016_iceXVII,Wang_PRL_2020_H2OH2_newphase, Carvalho_Crystals_2022_lowTC2, Mao_PNAS_2004_sII, Vos_PRL_1993_H2H2O_discovery, Vos_ChemPhysLett_1996_H2OH2_HP, Machida_JCP_2008_XRAY_ic, Hirai_JCP_2012_PT_C2, Bove_PhilTransRSoc_2019_filledice, Kuzovnikov_JPCC_2019_H2OH2_pd} and hydrogen molecules occupy defined positions inside the ice-like channels. 
In the filled ice C$_1$ phase, which is constituted by an ice II-like host frame and has a 1:6 hydrogen to water molecular ratio \cite{Wang_PRL_2020_H2OH2_newphase, Carvalho_Crystals_2022_lowTC2, Vos_PRL_1993_H2H2O_discovery}, the first rotational energy level of the confined H$_2$ was found (by inelastic neutron scattering experiments at high pressure and low temperature) to be only marginally  different from that of a free H$_2$ rotor \cite{Ranieri_JPCC_2019_quantumH2}. 
Similar results \cite{Del_Rosso_Phys_Rev_Mat_2017_C0} were observed for the quantum dynamics of H$_2$ confined in the C$_0$ structure of hydrogen hydrate, where the host water lattice is not based on any stable
ice phase \cite{Strobel_JACS_2016_C0,Amos_JPCL_2017_C0,DelRosso_NatComm_2016_iceXVII}.
Among the diverse filled ice phases, the C$_2$ phase (stable above $\sim$2.5 GPa) stands out for its distinctive characteristics arising from the interplay between hydrogen and water molecules, which form two identical and interpenetrated ice I$c$-like (or diamond-like) sublattices with a 1:1 hydrogen to water molecular ratio. As a result, hydrogen molecules are very densely packed in this structure and the hydrogen--water distances are very short \cite{Vos_PRL_1993_H2H2O_discovery, Vos_ChemPhysLett_1996_H2OH2_HP, Machida_JCP_2008_XRAY_ic, Hirai_JCP_2012_PT_C2, Komatzu_nat_comm_2020}.
One of the remarkable features of the  C$_2$ phase is indeed the profound influence of the surrounding water molecules on the quantum dynamics of the encapsulated hydrogen molecules. Unlike in other phases, in the C$_2$ phase, the crystal field imposed by the water molecules is expected to strongly affect the quantum rotational dynamics of the H$_2$ molecules. 
To our knowledge, the quantum mechanical behavior of molecular hydrogen within the C$_2$ lattice has not been explored yet.
Above $\sim$40 GPa, the C$_2$ phase transforms into the C$_3$ phase upon laser heating \cite{Qian_SciRep_2014_C3_pred, Ranieri_PNAS_2023}, which has an identical cubic water sublattice but is even richer in H$_2$ (2:1 molecular ratio), resulting in significant unit cell expansion and consequent larger hydrogen--water distances. 
Therefore, the C$_2$  phase represents the most extreme case of confinement for hydrogen molecules in a water frame and thus provides an ideal benchmark to study the quantum motion of hydrogen in strong confinement.


In this work, we employed state-of-the-art computational and experimental techniques to fully characterize the rotational quantum dynamics of the H$_2$ molecules in the C$_2$ phase at low temperatures. In particular, we solve the 3D Schr\"{o}dinger's equation for a hydrogen molecule embedded in the effective crystal field of the C$_2$ phase: the H$_2$ effective potential at finite temperature computed within \emph{ab initio} molecular dynamics (AIMD) accounting for van der Waals corrections to the exchange-correlation functional. Our calculations reveal that the H$_2$O--H$_2$ interaction is such to induce a giant splitting of $\pm 3.2$ meV in the $l$=1 rotational eigenenergies.
To validate our theoretical predictions and further probe the quantum rotational dynamics of H$_2$ within the C$_2$ phase, we employ inelastic neutron scattering (INS), a powerful experimental technique capable of providing detailed insights into the energy levels of hydrogen molecules. Through pioneering INS experimental measurements conducted at 6 GPa and low temperature (4 K $<$ $T$ $<$ 85 K), we corroborate our theoretical findings, thereby affirming the accuracy of our predictions regarding the quantum rotational dynamics of H$_2$ within the C$_2$ phase.

Our results not only deepen the understanding of the fundamental physics governing hydrogen hydrates and shed light on the hydrogen--water interaction in dense systems, but also carry broader implications for molecular hydrogen under extreme conditions. Furthermore, the observed heightened induced-dipole interactions within the C$_2$ phase underscore the significance of hydrogen hydrates in diverse fields ranging from materials science to astrochemistry.



\begin{figure}[ht]
    \centering
	\includegraphics[width=1.0\columnwidth]{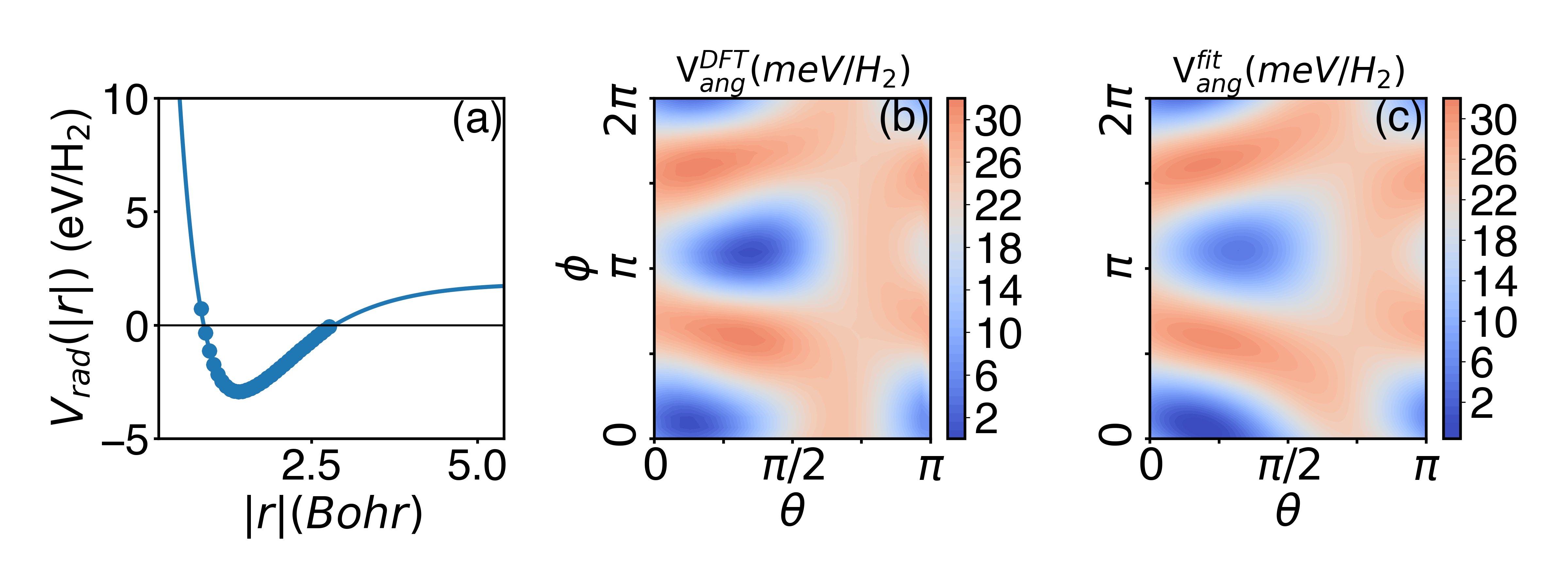}
	\caption{Panel (a): Radial part of the potential ($V_{rad}$), DFT-calculated data points are shown as blue circles, the fit with the Morse potential is shown as a blue line. The energy zero is taken as the last calculated point. Panels (b) and (c) show the calculated and fitted angular part of the potential ($V_{ang}(\theta, \phi)$), respectively, at a temperature $T$ = 10 K. See Supplementary Material \cite{SM} for further details.}
	\label{mainfig:fitpotential}
\end{figure}

The H$_2$ molecule is the textbook example of a quantum rigid rotor, with rotational peaks equally spaced and separated by 2$B$ = $\hbar^2 / I$, where $I$ is the moment of inertia $\mu R^2 = \frac{m R^2}{2}$, with $\mu$ the reduced mass, $m$ the mass of H$_2$, and $R$ the H--H distance.  
The Schr\"{o}dinger's equation extends to the case of an external potential $V_{ext}$ (Cartesian coordinates, atomic units): 
\begin{equation}
    \left(-\frac{\nabla^2}{2 \mu}+V_{ext}(\vec{r}) \right)\psi_{n}(\vec{r}) = \epsilon_{n}\psi_{n}(\vec{r})
    \label{eq:schrodinger}
\end{equation}
For a H$_2$ molecule in the C$_2$ crystal, the induced-dipole--dipole interactions between H$_2$ and the water sublattice can be modeled precisely as such an effective potential $V_{ext}$. We neglect quantum fluctuations in the hosting lattice. Within this approximation, at \SI{0} {\kelvin}, $V_{ext}$ is the total energy of the crystal at various hydrogen orientations. Temperature can also be accounted for by replacing $V_{ext}$ with the free energy landscape averaging over different snapshots from molecular dynamics at the chosen temperature (see Supplementary Material (SM) for further details \cite{SM}). The calculations were performed using the DFT-relaxed structure of C$_2$ having the correct experimental volume (see Tab. S2 in SM \cite{SM} for full structural details). The result is the effective potential shown in Fig. \ref{mainfig:fitpotential}, separated into its radial ($|r|$) and angular ($\theta, \phi$) components. The radial component is very well fitted by a Morse potential, while the angular part is characterized by a complex landscape well reproduced by a combination of sines and cosines (see SM \cite{SM}).

\begin{figure}[!tb]
	\includegraphics[width=0.99\columnwidth]{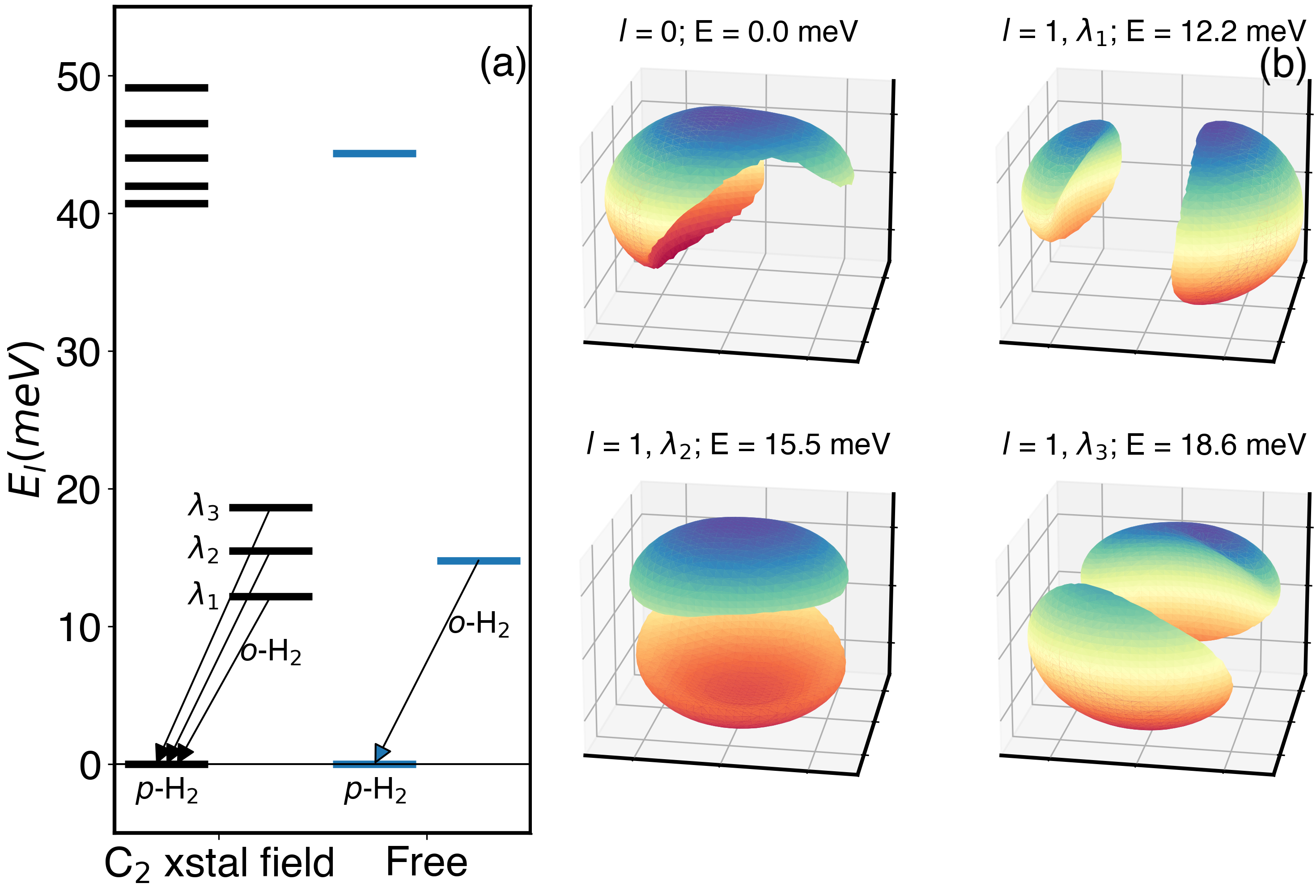}
	\caption{Panel (a): Calculated energy levels for the quantum rotations of the H$_2$ molecule in the C$_2$ phase (black) and for the free rotor (blue). The eigenenergies are divided into \textit{even}-$l$ (\textit{p}-H$_2$) and \textit{odd}-$l$ ones (\textit{o}-H$_2$). Arrows indicate the transition from ortho-H$_2$ ($l$=1) to para-H$_2$ ($l$=0). Panel (b): Isosurfaces showing the eigenfunctions of the ground state and the first three excited states of the H$_2$ molecule in the C$_2$ crystal field (isolevel: $7e-5$), the color indicates the $z$ axis and is used as a visual aid. The total angular momentum $l$, index $\lambda$, and eigenenergies are given.}
	\label{fig:C2_enlevels}
\end{figure}

The Sch\"{o}dinger's equation \ref{eq:schrodinger} with the calculated $V_{ext}$ was solved numerically using the Lanczos algorithm in shift-invert mode to obtain the lowest-energy eigenvalues and eigenfunctions. In Fig. \ref{fig:C2_enlevels}(a), we show the energy states compared to the free rotor. We remark that H$_2$ molecules being constituted of two identical fermionic hydrogens form two spin isomers, namely \textit{para}-hydrogen ($p$-H$_2$) with nuclear spin number I=0 and \textit{ortho}-hydrogen ($o$-H$_2$) with nuclear spin number I=1. Due to total wavefunction symmetry constraints, $p$-H$_2$ can only be associated with even rotational quantum number states ($l$=0,2,4...) and $o$-H$_2$ with odd states ($l$=1,3...). To benchmark the solver, Fig. \ref{fig:C2_enlevels}(a) shows that our method is consistent with the analytic result in the absence of $V_{ext}$. When the host lattice is accounted for,  the magnetic quantum number $m$ degeneracy is broken with a splitting of $\pm$3.2 meV for $l=1$. 

In Fig. \ref{fig:C2_enlevels}(b), we show the wavefunctions corresponding to the ground state and the first three excited levels. 
The host crystal field breaks the rotational symmetry of the molecule, and both $l$ and $m$ are no longer good quantum numbers. However, the low energy states still keep a dominant $l$ value that can be used to group states, while the $m$ states with a similar $l$ are mixed to form oriented orbitals that we number with an index $\lambda$ (see Fig.~\ref{fig:C2_enlevels}b and Sect. S2B of SM \cite{SM} for more details). It is easier to interpret the wavefunctions in comparison with the spherical harmonics, i.e., the result for the unperturbed rotor (shown in Fig. S6): the ground-state wavefunction is shaped like a hollow hemisphere, i.e., localized around $|\vec{r}| \sim R$. Unlike the unperturbed rotor (where the wavefunction is a perfect sphere), this wavefunction exhibits a broken angular symmetry in the direction of the minimum of the potential. The first excited state wavefunction is oriented roughly in the same direction as the ground state, while the other two point in orthogonal directions, like the $p_x$, $p_y$, and $p_z$ angular orbitals of the hydrogen atom.

\begin{figure}[!b]
	\includegraphics[width=\columnwidth]{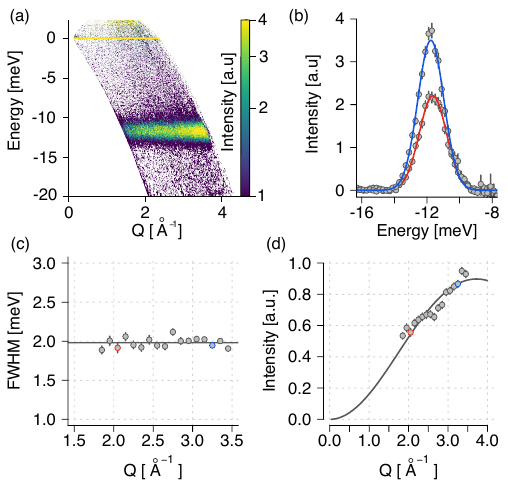}
	\caption{INS experimental data acquired on IN5 at $P$=6.0 GPa and $T$=4.4 K. Panel (a): $S(Q,\omega)$ after data reduction. A strong inelastic feature is observed at about -11.7 meV. Panel (b): $Q$ binned spectra and their Gaussian fits for $Q$ = 2.05 \AA$^{-1}$ (red line) and $Q$ = 3.25 \AA$^{-1}$ (blue line). Panel (c): FWHM of the Gaussian fits to the $Q$-sliced INS spectra as a function of $Q$. Panel (d): Integrated intensity of the Gaussian fits to the $Q$-sliced INS spectra as a function of $Q$ and its best fit (see Supplementary Material \cite{SM} for further details).}
	\label{fig:sqomega}
\end{figure}

Transitions between quantum rotational levels can be directly measured with inelastic neutron scattering. In particular, as transitions between the ground state ($l$=0) and the first excited state ($l$=1) involve changes in the nuclear spin, INS is the only experimental technique able to directly measure the fundamental rotational transition of molecular hydrogen. In addition, thanks to the long-lived metastable nature of the \textit{ortho} state \cite{ortho_para}, the \textit{ortho}-\textit{para} transition can be measured in the neutron energy gain side of the dynamical structure factor even at low temperature (see paragraph S5 E in \cite{SM}) \cite{Horsewill_PRB_2012_H2_inelastic}. Previous studies focusing on hydrogen molecules trapped in molecular cages of different sizes showed that rotational transitions can combine with low-energy rattling modes \cite{Colognesi_JPCA_2013_H2_sII,Ranieri_JPCC_2019_quantumH2}, leading to many possible combinations. In the C$_2$ hydrate, however, rattling modes are out of the probed energy window due to the short average H$_2$--H$_2$O distances, leading to a much simpler INS spectrum, as shown in Fig. \ref{fig:sqomega}(a). 

Two INS experiments were performed on the IN5 time-of-flight spectrometer at ILL in Grenoble, France (raw data available at \cite{exp1} and \cite{exp2}). The H$_2$:D$_2$O sample was loaded cryogenically in a Paris-Edinburgh press (see SM \cite{SM} for further details on the sample preparation and environment). In the first run, the sample was compressed to 6.0 GPa (see Fig. S9) and cooled down to 4.4 K; in the second, it was cooled to 9 K and the incoherent dynamical structure factor $S(Q,\omega)$ was measured using an incident wavelength of 4.8 \AA, thus mainly measuring in the neutron energy gain side. The results from the first run are shown in Fig. \ref{fig:sqomega}, while those from the second run are shown in SM (Figure S11) \cite{SM}. 

\begin{figure}[!t]
	\includegraphics[width=\columnwidth]{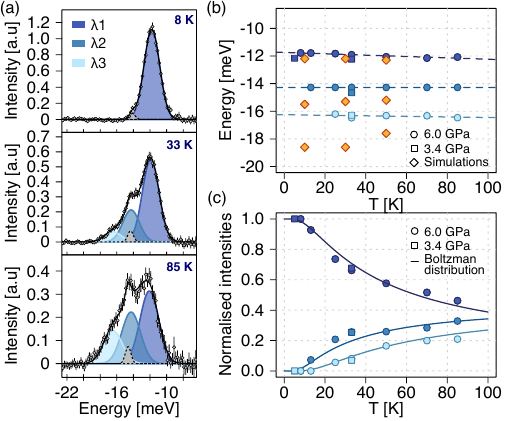}
	\caption{$T$ dependence of the INS experimental results. Panel (a): INS spectra at 6 GPa and 8 K, 33 K, 85 K, compared to their best fits (blue areas). The gray shaded area is the peak from the pure solid hydrogen in the sample. Panel (b): Energies of the three rotational peaks at 6.0 and 3.4 GPa (blue circles and squares respectively) as a function of temperature compared with numerical results (yellow diamonds). Panel (c): Normalised intensities of the three peaks at 6.0 and 3.4 GPa (blue circles and squares respectively) compared to the statistical weights of the initial states calculated using the Boltzmann distribution. Plots with error bars are reported in Fig. S13 in Supplementary Material \cite{SM}. The different blue shades are consistently associated to the transitions from the $\lambda_1$, $\lambda_2$, $\lambda_3$ levels to the ground state as shown in the legend.}
	\label{fig:Tdependence}
\end{figure}

In both runs, we observed the first \textit{ortho-para} ($l$=1, $\lambda_1$ to $l$=0) transition of the hydrogen molecules confined in the C$_2$ phase to have an energy of 11.7 meV, i.e. about 3 meV lower compared to the ortho-para transition of free hydrogen (14.6 meV). The rotational nature of the excitation peak observed in the $S(Q,\omega)$ is confirmed from the exchanged wavevector $Q$  analysis of both peak width and intensity. After $Q$-slicing the $S(Q,\omega)$, the spectra (Figure \ref{fig:sqomega} panel (b)) can be fitted with Gaussian functions, whose FWHM are found to be constant with $Q$ (panel (c) of Fig. \ref{fig:sqomega}) and whose intensities are well described by a squared first-order spherical Bessel function (panel (d) of Fig. \ref{fig:sqomega}). Both of these trends are indicative of a rotational nature for the excitation.  As the temperature was increased up to 85 K, we observed a progressive population of the higher energy levels ($l$=1, $\lambda_2$ and $l$=1, $\lambda_3$) and the consequent associated transition to the ground state (see Fig. S10 for a schematic representation of the observed transitions). The occupation of the levels is in good agreement with the Boltzmann distribution and the splitting of the $l=1$ level is of $\pm$ 2.2 meV, 30\% smaller than in our calculations (yellow diamonds in Fig. \ref{fig:Tdependence} panel (b)). Examples of INS spectra (obtained by integrating the $S(Q,\omega)$ over the whole measured $Q$ range), compared with their best fits, and fit results are shown in Fig. \ref{fig:Tdependence}. The $Q$ analysis of the data shows a consistent behavior also at the highest investigated temperatures (see Fig. S15). The sample was then decompressed to 3.4 GPa at 85 K and measured at 5 K and 33 K (squares in Fig. \ref{fig:Tdependence} panel (b), spectra are reported in S16), where we observed a 7.5\% reduction of the energy splitting. In addition, we observed that the rotational transitions show an intrinsic broadening much larger than our experimental resolution (0.8 meV), which can be attributed to structural defects formed upon compression (see Fig. S17).

Overall, we find the calculated and experimental results for the energy levels in excellent agreement in terms of both the absolute values and the energy splitting (see Fig. \ref{fig:Tdependence} panel (b)). 
The largest discrepancy is for the high-energy mode, which decreases with increasing temperature. This is expected as the temperature dependency of our model comes from the thermal activation of the phonons of the host water frame, altering the effective potential on the H$_2$ molecule. At very low temperatures, the zero-point motion of the ice skeleton kicks in, which is neglected by our classical MD sampling. Indeed, as temperature increases, the classical sampling of the ice phonons improves as the agreement between simulations and experimental data. 


In summary, we studied the quantum dynamics of the H$_2$ molecule in the C$_2$ filled ice phase at high pressure and low temperature. The overall excellent agreement between our calculations and the experimental results offers a simple interpretation of the data. In the C$_2$ phase, the H$_2$ and water interact quite strongly, leading to a strong angular anisotropy in the effective crystalline potential felt by the hydrogen molecule. This removes the angular degeneracy with respect to the magnetic quantum number $m$, resulting in an energy splitting (for the triplet $l$=1) of $\pm$3.2 meV in our calculations and $\pm$ 2.2 meV experimentally.
To the best of our knowledge, such a large splitting in the energy levels of a quantum rotor in a water-based clathrate is unprecedented. As the C$_2$ hydrate is stable (or metastable) over a wide pressure range spanning from nearly ambient pressure (at low temperature) \cite{Mao_PNAS_2004_sII,Komatzu_nat_comm_2020} to pressures above 80 GPa at least \cite{Ranieri_PNAS_2023}, it represents an ideal playground to further explore low-energy physics of quantum rotors in the future.
\\
\\
We acknowledge the Institut Laue-Langevin (ILL) for providing beamtime on IN5 and the assistance from Claude Payre with the high-pressure setup. We thank Leon Andriambariarijaona and Alasdair Nichols for their help during the experiments on IN5, and Thomas C. Hansen for technical assistance during the experiment on D20. We thank Lorenzo Ulivi, Leonardo del Rosso, and Milva Celli for providing the sample. S.D.C. acknowledges computational resources from CINECA, proj. IsC90-HTS-TECH and IsC99-ACME-C, the Vienna Scientific Cluster, proj. 71754 "TEST". L.E.B. acknowledges the financial support by the European Union - NextGenerationEU (PRIN N. F2022NRBLPT), the ANR-23-CE30-0034 EXOTIC-ICE, and the Swiss National Fund (FNS) under Grant No. 212889. 

%

\end{document}